\documentclass{article}
\usepackage{amsfonts}
\usepackage{amsmath}
\usepackage{amssymb}

\input{tcilatex}
\begin{document}

\title{Peculiarities of massive vector mesons and their zero mass limits\\
{\small published in Eur.Phys. J. C (2015) 75:365}\\
{\small In memory of Raymond Stora}}
\author{Bert Schroer \\
CBPF, Rua Dr. Xavier Sigaud 150, \\
22290-180 Rio de Janeiro, Brazil\\
permanent address: Institut f\"{u}r Theoretische Physik\\
FU-Berlin, Arnimallee 14, 14195 Berlin, Germany}
\date{June 2015}
\maketitle

\begin{abstract}
Massive QED, in contrast with its massless counterpart, possesses two \
conserved charges; one is a screened (vanishing) Maxwell charge which is
directly associated with the massive vector mesons through the identically
conserved Maxwell current, while the presence of a particle-antiparticle
counting charge depends on the matter . A somewhat peculiar situation arises
for couplings of Hermitian matter fields to massive vector potentials; in
that case the only current is the screened Maxwell current and the coupling
disappears in the massless limit.

In case of selfinteracting massive vector mesons the situation becomes even
more peculiar in that the usually renormalizability guaranteeing validity of
the first order power-counting criterion breaks down in second order and
requires the compensatory presence of additional Hermitian H-fields.

Some aspect of these observation have already been noticed in the BRST gauge
theoretic formulation, but here we use a new setting based on string-local
vector mesons which is required by Hilbert space positivity ("off-shell
unitarity"). This new formulation explains why spontaneous symmetry breaking
cannot occur in the presence of higher spin $s\geq 1$~fields. The coupling
to $H$-fields induces Mexican hat like selfinteractions; they are not
imposed and bear no relation with spontaneous symmetry breaking; they are
rather consequences of the foundational causal localization properties
realized in a Hilbert space setting. In case of selfinteracting massive
vectormesons their presence is required in order to maintain the first order
power-counting restriction of renormalizability also in second order. The
presentation of the new Hilbert space setting for vector mesons which
replaces gauge theory and extends on-shell unitarity to its off-shell
counterpart is the main motivation for this work.

The new Hilbert space setting also shows that the second order Lie-algebra
structure of selfinteracting vector mesons is a consequence of the
principles of QFT and promises a deeper understanding of the origin of
confinement.
\end{abstract}

\section{Introduction}

The theoretical interest in massive vector mesons can be traced back to
Schwinger's conjecture \cite{Schwinger} which states that "massive QED"
leads to "charge screening". The analogy to the quantum mechanical theory of
superconductivity, where the long range vector potentials of
electromagnetism become short ranged, lends plausibility to Schwinger's
quantum field theoretical conjecture. This idea was made precise in a
theorem by Swieca \cite{Swieca} \cite{Sw} who showed that the Maxwell
current of a massive vector meson, independent of whether it interacts with
complex or Hermitian matter fields (in the sequel referred to a $H$-matter),
always leads to a vanishing charge of the identically conserved Maxwell
current.

The proof uses analytic properties of matrix elements (formfactors) of
identically conserved currents associated with a field strength tensor of a
massive vector meson. We will refer to this phenomenon as the
"Schwinger-Swieca screening". The global particle-antiparticle number
conservation remains unaffected by the vanishing of the Maxwell charge. In
case of a coupling to $H$-matter (the abelian Higgs model) there is only the
screened Maxwell charge.

Massive QED does not require the presence of $H$-particles and massive
vector mesons do not owe their mass to spontaneous symmetry breaking (SSB).
As will be shown in the present work, the mass of vector mesons bears no
intrinsic (physically meaningful) relation with a spontaneous symmetry
breaking "Higgs mechanism". As in the case of quantum mechanical
superconductivity, where long range vector potentials become short ranged 
\textit{without adding new degrees of freedom}, massive QED does not require
the presence of additional $H$-matter.

The situation changes in the presence of selfcouplings between massive
vector mesons as will be explained in the sequel. A necessary restriction on
first order couplings from the requirement of renormalizability is the
power-counting restriction for the short distance scale dimension of the
first order interaction density $d_{sd}^{int}\leq 4.$ In all known
renormalizable models, except those involving \textit{selfinteracting}
massive vector mesons, this bound secures the preservation of
renormalizability to all orders. This "golden rule" of renormalization
theory is satisfied in first order selfinteractions between massive vector
mesons, but is lost in second order; although there exists a unique first
order gauge invariant density for massive selfinteracting vectormesons with $%
d_{sd}^{int}=4,\ $the implementation of second order gauge invariance for
the S-matrix induces terms which violate the power-counting restriction.

It turns out that this second order violation can only be prevented by a
compensation in which an additional field couples to the vector mesons in
compliance with the first order power-counting restriction but a violation
in second order. This coupling must be such that the contributions from both
second order terms cancel. The charge neutrality of vector mesons requires
the new fields to be Hermitian; it turns out that the compensation can only
be achieved with scalar $H$-fields. Such a cancellation is reminiscent of
short distance improvements from compensations between different spin
components in supersymmetric interactions, but it is well-known that they
did not suffice in order to remain within the power-counting restriction.

However for the case at hand the compensation works; $A_{\mu }$
selfcouplings and $A$-$H~$coupling collaborate in such a way that
renormalizability is preserved in second order. This scenario was first
elaborated in the setting of BRST gauge theory by Scharf \cite{Scharf} and
collaborators \cite{Aste}, for a more recent account see also \cite{BDSV}.
They start from the perturbative formal representation of the S-matrix%
\footnote{%
The physical S-matrix is the adiabatic limit of the Bogoliubov genertaing
operator S-functional.} in terms of the adiabatic limit of time-ordered
products of the first order interaction density and impose BRST gauge
invariance in the form $\mathfrak{s}S=0,$ where $\mathfrak{s}$ denotes the
nilpotent BRST s-operation. \ 

This turns out to be very restrictive; in case of only one $H~$fields all
couplings are determined. As expected, in the limit of zero mass vector
mesons the $H$ decouples from the $A_{\mu }$ and becomes a free field,
whereas the massless selfinteracting vectormesons take the form of a
(massless) Yang-Mills (YM) interaction.

Besides the BRST gauge formalism, needed for the definition of the gauge
invariant S-matrix, no other property has been used to derive this result.
In particular there is no reference to SSB which \textit{starts} with the
Mexican hat potential, whereas in the present setting it is \textit{induced}
from the imposition of second order gauge invariance.

What is meant by "induction" is best explained in the simpler abelian Higgs
model. The starting point of any perturbation theory is a set of
interaction-defining free fields and a first order coupling strength; in the
present case these are 3 parameters, namely the masses of the vector meson
and the $H,$ and the first order coupling strength $g$ in $gA_{\mu }A^{\mu
}H.$ Among the terms, which the imposition of gauge invariance (BRST $%
\mathfrak{s}$-invariance) on the S-matrix induces in second order, one finds
the tri- and quadri-linear terms which can be written in the form of a
Mexican hat potential which only depends on $g$ and ratios of the two masses 
\cite{Scharf}. This kind of induction, which generates new interactions
without enlarging the number of coupling parameters, is a consequence of
gauge invariance of $S.$

The construction of the (abelian) Higgs model in terms of a SSB prescription
is very different. It starts with the Lagrangian of two-parametric scalar
QED and breaks gauge invariance in terms of a shift in field space. A
subsequent gauge transformation converts the resulting expression, apart
from the unphysical parametrization, into the same expression as that
obtained by imposing gauge invariance. An easy test, which shows that the
shift prescription has no relation to a physical SSB, is to pass to the
limit of a vanishing shift parameter. In a genuine SSB the model would
return to its unbroken form which was 2-parametric scalar QED. This does not
happen, the physically meaningless application of a gauge transformation
after breaking of gauge invariance prevents this return.

A more profound argument results from the observation that the Maxwell
current of the Higgs model, as any Maxwell current of a massive vectormeson,
leads to a screened charge $Q=0$ whereas the conserved current of a genuine
SSB model implies $Q=\infty .$ \textit{A conserved current whose "would be"
symmetry-generating charge diverges is actually the definition of a SSB, }%
whereas the shift in field space is a device which prepares such a current.%
\textit{\ }

It is well known that such situations can be generated from quartic
interactions between scalar particles which are invariant under the action
of compact$~$groups. The shift in field space is a convenient tool to
construct a first order interaction whose renormalized perturbation theory
preserves current conservation but leads to the divergence of some of the
charges. It does not create masses but rather prepares the kind of special
interaction between massive and massless ("Goldstone bosons") free fields
which maintains current conservation but causes the long distance divergence
of some charges.

This construction is limited to interactions involving low spins $s<1.$ The
coupling of vectormesons to scalar fields prevent the latter from causing a
SSB. The usual argument is that the scalar particles become gauge-dependent
and local gauge invariance is not really a physical symmetry which can be
broken. This argument involves formal steps with unclear physical content;
gauge "symmetries" are not physical symmetries and it is not clear what one
is breaking.

A less formal more physical argument is to look directly at the gauge
invariant conserved current which the coupling to the vectormeson has
transformed into an identically conserved Maxwell current. The charge of
Maxwell currents of massive vector mesons vanish as a consequence of the
Schwinger-Swieca charge screening; this is a gauge-invariant phenomenon (it
involves only gauge invariant observables). A screened charge $Q=0$ is very
different from a SSB charge $Q=\infty ;$ between the two is the nontrivial
charge $Q<\infty $ which is the generator of a symmetry. As soon as higher
spin fields couple to scalar fields, the latter are prevented to undergo
SSB; the scalar particles have to follow the more restrictive nature of
higher spin interactions places the vector mesons into the driver's seat. In
this way those scalar fields which couple to $s=1$ fields are prevented from
transforming into their SSB mode.

In the string-local field (SLF) Hilbert spaces setting, whose presentation
is the main purpose of the present paper, the interaction with string-local
vector mesons converts the scalar fields into interacting string-local
fields. In this setting there is simply no gauge symmetry which could
undergo spontaneous breaking; rather all point-local fields which couple to
string-local vector potentials loose their point-local nature and become
string-local. Such fields are not local observables. But in contrast to
their gauge theoretic counterparts they are physical since they act in a
Hilbert space.

The Higgs mechanism leads to the correct coupling of $H~$with massive vector
mesons but it fails to reveal the \textit{raison d'\^{e}tre} of the $H$%
-field which has nothing to do with symmetry breaking and mass generation.
The vector mesons in massive QED does not need the presence of a
mass-generating $H,~$it is a renormalizable theory by itself. As pointed out
before, the situation changes radically in the presence of selfinteracting
massive vector mesons. In that case the theoretical reason for its existence
is the second order preservation of renormalization. As a result of the
connection between renormalizability and causal localizability, the $H$
plays a fundamental role for the consistency of the $W^{\pm },Z$
interactions in the Standard Model.

The preservation of second order renormalizability with the help of an $H$%
-particle is vaguely reminiscent of the introduction of vector mesons to
convert the first order nonrenormalizable 4-Fermi interaction into a
renormalizable coupling, except that a second order compensation mechanism
is much more sophisticated. The fundamental nature of the LHC discovery is
in no way affected by this new role assigned to the $H,$ but in particle
theory it is important to distinguish between prescriptions for construction
of models and their intrinsic properties. \textit{\ }

It is an interesting question why this was not seen in the standard BRST
formulation in terms of Feynman rules. The answer is that in an off-shell
gauge formalism one usually presents the perturbative rules, but one does
not focus on the explicit construction of a gauge-invariant on-shell unitary
S-matrix. In \cite{Scharf} \cite{Aste} the BRST formalism was especially
adjusted to that problem by what the authors called the "causal gauge
invariance" (CGI) setting which is based on the Epstein-Glaser operator
formulation. This leads to a rather clear distinction between quasiclassical
pictures of symmetry breaking and gauge-induced $H$-selfinteractions. \ 

This CGI setting permits a direct perturbative construction of a unitary 
\textit{gauge invariant} S-matrix for a $gA\cdot AH$ coupling of a massive
vector meson with a Hermitian field; in this way it highlights the second
order induction mechanism which leads to the Mexican hat $H$%
~selfinteraction. It bears no relation with generating masses of vector
mesons; whereas massive QED does not need the presence of $H$ fields in
order to generate the mass of the vector meson, their presence is necessary
in all interaction involving selfinteracting massive vector mesons and the
CGI formulation reveals the correct reasons.

Interactions in QFT cannot generate masses in any material sense; masses of
interacting-defining \textit{elementary} fields (i.e. those in terms of
which the first order interaction is defined) must be put in, and higher
order renormalization theory preserves them. What one expects from the
theory is that the mass of possible bound states, which correspond to
composites of the elementary fields, can be computed within the theory; but
this does not seem to be possible within a perturbative setting.

The main concern in the present work is to replace the "ghostly" gauge
theory by a new ghost-free Hilbert space formulation of interactions for $%
s=1 $ fields.

Gauge theory is not a substitute of a Hilbert space formulation, but it is a
rather successful placeholder. The awareness about its makeshift status was
much stronger in the past than it is now. There were several valiant
attempts to avoid the use of indefinite metric by Mandelstam \cite{Man} and
also DeWitt \cite{DeWitt}; their failure also revealed that a description
which is consistent with positivity ("off-shell unitarity") requires major
new conceptual investments beyond what was known at that time.

As a result of impressive observational successes of gauge theory in its
applications to the Standard Model, this problem moved gradually into the
background. Physicists of the older generation sometimes express their
surprise that the conceptual incomplete gauge theory works much better than
expected; occasionally they think of this success in terms of as a results
of unmerited luck (I thank Raymond Stora for sharing his views on this
problem).

The new Hilbert space setting reveals what can be correctly described within
the perturbative gauge theory setting and what problems remain outside its
physical range. The vacuum sector, generated by the application of gauge
invariant observables to the vacuum state, and the gauge invariant S-matrix
in the presence of a mass gap are within its physical range, whereas the
construction of causally localized fields and their asymptotically related
particles states remain outside.

This is a serious limitation in particular in zero mass limits when the
construction of the gauge invariant S-matrix fails and one has to take
recourse to calculational recipes without understanding their connection
with the foundational spacetime localization properties. Such problems
related to the long-distance behavior of fields (infrared problems in
momentum space) are outside the range of gauge theory.

Local gauge symmetry is a well-defined concept in classical field theory,
but it clashes with the Hilbert space positivity of quantum theory. Instead
of representing a physical symmetry, it is a formal device for extracting a
physical subtheory. This is a consequence of the fact that locality in a
Krein space, in contrast to Einstein causality in Hilbert space, has only a
formal but no physical motivation. Its become particularly annoying in the
massless limit, when scattering amplitudes suffer from infrared divergencies
and one is forced to describe collisions in terms of momentum space
prescriptions instead of spacetime localization properties of physical
fields. Such problems can only be solved in a Hilbert space setting; the
operators in QFT which corresponds to long range (Coulomb) potentials in
quantum mechanics are string-local fields \cite{B}.

The new SLF theory should explain why selfinteracting vector mesons lead to
a Lie algebra structure with only one coupling strength. In the BRST gauge
setting this arises as a consistency condition of the formalism \cite{Scharf}%
. This is not surprising since the BRST formalism is the result of an
adaptation of classical gauge theory to the exigencies of QFT \footnote{%
I thank Raymond Stora for his critical remarks about such derivations.}. The
physically more relevant question is whether the Lie algebra structure can
be derived solely from the causal localization principles of QFT in a
Hilbert space setting without referring to the classical mathematics of
fibre bundles. In section 6 it will be shown that this is indeed the case.

A somewhat surprising result is that the use of covariant string-local
potentials permits a simple and amusing description of topological effects
which are present in the zero mass limit as the breakdown of Haag duality
and the closely related Aharonov-Bohm effect (section 3).

The interest in string-local fields, as they are used in the present work,
started with the solution of an old problem which goes back to Wigner. In
his famous 1939 paper, which contains the classification of all positive
energy representations of the Poincar\'{e} group, Wigner found besides the
massive and the zero mass finite helicity representation a third massless
representation class which he referred to as "infinite spin". Whereas there
was no problem to associate point-local covariant fields with the first two
classes, the problem of a field theoretic description of the third class
remained for a long time open. In \cite{Y} it was shown that this class
cannot be described in terms of point-local Wightman fields. Using methods
of modular localization\footnote{%
Modular localization is an intrinsic formulation of causal localization
which does to rely on the use of particular field "coordinatizations".}
Brunetti, Guido and Longo \cite{BGL} showed that these representation permit
causal localization in arbitrary narrow space-like cone regions in Minkowski
space (for historical remarks on modular localization see \cite{EPJH} \cite%
{SHPMP}).

This suggested that it should be possible to associate covariant fields
localized on space-like semi-infinite strings (the "core" of arbitrary
narrow space-like cones) with such a situation; such fields were then
explicitly constructed in \cite{MSY}. In that work arguments were given
which suggested that such models do not admit composite point-local fields;
they were significantly extended in \cite{Koe}.\ An elegant proof of their
absence was recently given in \cite{LMR}.

In \cite{MSY} also string-local fields for the massive and zero mass finite
helicity Wigner representations were constructed in the hope that they could
be useful in perturbation theory and may lead to a ghost-free Hilbert space
formulation of $s=1$ interactions which replaces gauge theory. These ideas
were further pursued in \cite{Bros} \cite{Rio} \cite{charge} \cite{SHPMP} 
\cite{FOOP} \cite{sigma}. The present work is a continuation of these ideas
in the context of explicit second order calculations in the presence of
string-local massive vector potentials. The presentation of the general
formalism in the presence of string-crossings will be contained a
forthcoming paper by Jens Mund; further details about applications to
interacting string-local fields will be addressed in joint work by Mund and
the present author.

It had been known for a long time \cite{Bu-Fr2} \cite{Haag} that within the
setting of algebraic QFT the field-particle relation in the presence of a
mass gap can always be described in terms of operators localized in
arbitrary narrow space-like cones (whose cores are semi-infinite space-like
strings). It is interesting to note that the ideas which led to these
results arose from a previous publication of the authors in which they
removed a remaining loophole in the proof of Swieca`s screening theorem \cite%
{Bu-Fr1}.

The approach presented in this paper may be seen as an adaptation of those
structural results to the requirements of renormalized perturbation theory
in terms of string-local fields.

This work is organized as follows.

The next section presents conceptual aspects of the Hilbert space setting
and compares them with the formalism of operator gauge theory.

The third section contains a simple but somewhat surprising application of
free string-local fields.

The problem of their role in interactions is taken up in section 4.

Section 5 presents second order perturbative results which includes in
particular the interaction of massive vector mesons with Hermitian matter
(the Higgs model).

In section 6 it is shown that the Lie-algebra structure of selfinteracting
vector-mesons is a consequence of causal localization in Hilbert space.

The concluding remarks present an outlook about what can be expected from
this new setting concerning unsolved problems of infrared divergencies such
as "infraparticles" in QED and confinement in QCD.

\section{Formal analogies and conceptual differences between CGI and SSB}

For the convenience of the reader we start with a compilation of formulas as
they are used in the CGI formulation of BRST gauge theory for the
construction of the gauge invariant S-matrix \cite{Scharf}.%
\begin{eqnarray}
&&\mathfrak{s}A_{\mu }^{K}=\partial _{\mu }u^{K},~\mathfrak{s}\phi
^{K}=u^{K},~\mathfrak{s}\hat{u}^{K}=-(\partial A^{K}+m^{2}\phi ^{K})
\label{Krein} \\
&&\mathfrak{s}B:=i[Q,B],~Q=\int d^{3}x(\partial ^{\nu }A_{\nu
}^{K}+m^{2}\phi )\overleftrightarrow{\partial }_{0}u  \notag
\end{eqnarray}%
Here the superscript $K$ refers to the Krein space in which these operators
are realized, $Q$ is the so-called ghost charge whose properties ensure the
nilpotency ($\mathfrak{s}^{2}=0$) of the BRS $\mathfrak{s}$-operation. The $%
A_{\mu }^{K}$ is a massive vector meson in the Feynman gauge and $\phi ~$is
a free scalar field of the same mass but with a two-point function of
opposite sign (a kind of negative metric St\"{u}ckelberg field); these two
fields carry the indefinite metric which requires to replace the Hilbert
space by a Krein space (iteratively created by iterative application of the
operators to the vacuum).\ The "ghosts" $u,\hat{u}$ are free "scalar
fermions" whose presence is necessary in order to recover the vacuum sector
of the local observables acting in an Hilbert space and the unitary S-matrix
in the form of $\mathfrak{s}$-invariant operators.

These rules in terms of free fields suffice for the construction of the
gauge-invariant S-matrix; the extension to (gauge-variant) interacting
fields follows similar formal rules as those for interacting fields with
lower spin $s<1$. The $Q$-charge and the $\mathfrak{s}$ participate in the
perturbation theory of interacting fields.

The BRST formalism is a pure perturbative tool; structural properties (TCP,
spin\&statistics,..) as well as the physical causal locality properties of
which they are consequences require the Hilbert space positivity. The tools
one needs for nonperturbative constructions and the derivation of structural
theorems (Schwartz inequality,..) are not available in a Krein space
setting. Quantum gauge theory is limited to the combinatorical manipulations
of perturbation theory.

It should not come as a surprise that the ghost formalism (the $\mathfrak{s}$%
-cohomology), unlike the later SLF Hilbert space formulation (differential
forms on the $d=1+2~$unit de Sitter space of space-like directions), has no
relation with spacetime. The BRST rules were not derived from localization
principles of QFT but they were found in the course of trying to recover
unitarity of the S-matrix ("on-shell unitarity") in a Krein space setting.
In order to arrive at the BRST operational formulation it needed several
improvements of the original unitarity arguments of 't Hooft-Veltman
(Faddeev-Poppov, Slavnov) in order to reach the formally elegant ghost
formalism of Becchi-Rouet-Stora and Tyutin.

Although the use of these prescriptions turned out to be essential for the
success of the Standard Model, their conceptual relation with the
foundational principles remained unclear. QFT is the realization of causal
localization in a Hilbert space; without positivity ("unitarity") there is
no probability interpretation and hence the relation with quantum theories
foundational property is lost in gauge theory; it can only be recovered in
special (gauge-invariant) situations. It needs to be emphasized that
classical gauge theory is not affected by these shortcomings since the
foundational Hilbert space structure is characteristic of quantum theory.
Hence it is not surprising that the SLF Hilbert space description is outside
Lagrangian quantization (but not outside perturbation theory).

As already mentioned in the introduction, massless $s\geq 1~$covariant
tensor potentials are necessarily string-local. Massive point-like
potentials exist, but as a result of their short distance dimension $%
d_{sd}^{s}(point)=s+1$ their interactions are nonrenormalizable since
interactions formed with them violate the power-counting limit $%
d_{sd}^{int}\leq 4.~$The fact that the smallest possible short distance
dimension of string-local fields is $d_{sd}^{s}(string)=1$ suggests that
there may be renormalizable string-local interactions. But the
power-counting limit is not the only restriction, physics demands the
existence of sufficiently many local observables generated by point-local
fields and the preservation of string-localization. Furthermore the S-matrix
in models in with a mass gap should be independent of the string directions $%
e;$ although fields may be string-local, the particles which they
interpolate remain those string-independent objects whose wave-function
spaces were classified by Wigner. In the following we will show how these
requirements can be met for massive vector mesons.

We start from a massive vector potential (the Proca field) and define its
associated string-local potential in terms of the Proca field strength

\begin{eqnarray}
F_{\mu \nu }(x) &=&\partial _{\mu }A_{\nu }^{P}(x)-\partial _{\nu }A_{\mu
}^{P}(x),~~A_{\mu }(x,e)=\int_{0}^{\infty }F_{\mu \nu }(x+\lambda e)e^{\nu
}d\lambda ,~  \label{def} \\
\phi (x,e) &=&\int_{0}^{\infty }A_{\mu }^{P}(x+\lambda e)e^{\mu }d\lambda
,~~e^{2}=-1  \notag
\end{eqnarray}%
Whereas the short distance dimension of the Proca potential and its field
strength is $d_{sd}^{P}=2$, the string-local vector potential and its scalar
"escort" $\phi $ have $d_{sd}=1.$ We could of course change the "population
density" of degrees of freedom on the semi-infinite line from $p(\lambda
)\equiv 1$ (as above) to any other smooth function which approaches $1$
asymptotically without leaving the local equivalence (Borchers) class; but
if we want in addition to uphold the linear relation which follows from (\ref%
{def}), 
\begin{equation}
A_{\mu }(x,e)=A_{\mu }^{P}(x)+\partial _{\mu }\phi (x,e)  \label{rel}
\end{equation}%
we must use the same $p(\lambda ).$ Formally this corresponds to the
possibility of gauge changes in (\ref{Krein}); $p(\lambda )~$changes
preserve relative localization and a fortiori do not change the particle
content in the presence of interactions.

In contrast to the "virtual" strings of anyons/plektons \cite{anyon} which,
similar to cuts in complex function theory, may be displaced as long as
crossings are prevented, the strings of string-local fields are "real".
Needless to mention that the string-local potentials are the only vector
potentials which permit a $m\rightarrow 0$ limit.

For the following it turns out to be convenient to express (\ref{def}) as
linear relations between $v$-intertwiners 
\begin{eqnarray}
&&A_{\mu }^{P}(x)=\frac{1}{(2\pi )^{3/2}}\int
(e^{ipx}\sum_{s_{3}=-1,0,1}v(p)_{\mu ,s_{3}}a^{\ast }(p,s_{3})+h.c.) \\
&&v_{\mu ,s_{3}}^{A}(p,e)=v_{\mu ,s_{3}}(p)+ip_{\mu }v_{s_{3}}^{\phi },\text{
}v_{s_{3}}^{\phi }(p)=\frac{iv_{s_{3}}\cdot e}{p\cdot e+i\varepsilon } 
\notag
\end{eqnarray}%
where the second line is the relation (\ref{rel}) rewritten as a linear
relation between the three intertwiners. Here $v~$refers to the intertwiner
between the 3-component unitary Wigner representation and the covariant
vector-components of the Proca potential. Using the differential form
calculus on the d=1+2 de Sitter space of space-like directions, we define
exact 1- and 2-forms 
\begin{eqnarray}
u &=&d_{e}\phi =\partial _{e_{\alpha }}\phi de^{\alpha },~v_{s_{3}}^{u}=i(%
\frac{v_{\alpha ,s_{3}}}{p\cdot e}-\frac{v\cdot e}{(p\cdot e)^{2}}p_{\alpha
})de^{\alpha }~  \label{forms} \\
\hat{u} &=&d_{e}(A_{a}de^{\alpha })=\int_{0}^{\infty }d\lambda F_{\alpha
\beta }(x+\lambda e)de^{\alpha }\wedge de^{\beta }~  \notag \\
v_{{}}^{\hat{u}} &=&p_{\alpha }\frac{v_{\beta ,s_{3}}}{p\cdot e}de^{\alpha
}\wedge de^{\beta }  \notag
\end{eqnarray}

They are field-valued differential forms with $d_{sd}=1~$which in a certain
sense represent the Hilbert space counterpart of the cohomological BRST
ghost formalism in Krein space. To maintain the simplicity of the covariant
formalism, the differential forms on the unit $d=1+2$ de Sitter space are
viewed as restrictions of the 4-dimensional directional $e$-formalism. The
notation $u,\hat{u}$ suggests that they may play the role of the
differential form analogue of the ghost in the BRST formalism.

These field-valued differential forms are natural extensions of string-local
fields; together with $A_{\mu }(x,e)$ and $\phi (x,e)~$they are the only
covariant string-local members of the linear part of the local equivalence
class of the free Proca field with $d_{sd}=1$. Their ghost counterparts are
important in the off-shell BRST formalism, but (apart from the appearance of 
$u^{(K)}$-terms in the $Q_{\mu }~$formalism) they disappear in the formula
for S.

The 2-point Wightman functions of string-local objects can be calculated
from their intertwiners or directly in terms of the line integral
representation of the string-local $A,\phi .$ fields. One obtains\footnote{%
More details will be contained in forthcoming work by Mund.}

\begin{eqnarray}
&&\left\langle A_{\mu }(x,e)A_{\nu }(x^{\prime },e^{\prime })\right\rangle =%
\frac{1}{(2\pi )^{3}}\int e^{-i(x-x^{\prime })p}M_{\mu \nu }^{A}(p)\frac{%
d^{3}p}{2p_{0}}  \label{2-point} \\
&&M_{\mu \mu ^{\prime }}^{A}(p;e,e^{\prime })=-g_{\mu \mu ^{\prime }}-\frac{%
p_{\mu }p_{\mu ^{\prime }}(e\cdot e^{\prime })}{(p\cdot e-i\varepsilon
)(p\cdot e^{\prime }+i\varepsilon )}+\frac{p_{\mu }e_{\mu ^{\prime }}}{%
(p\cdot e-i\varepsilon )}+\frac{p_{\mu }e_{\mu ^{\prime }}^{\prime }}{%
(p\cdot e^{\prime }+i\varepsilon )}  \notag
\end{eqnarray}%
and similar expressions for $M^{\phi }$ and mixed vacuum expectations $%
M_{\mu }^{A,\phi }.$ The occurrence of the latter (which vanish in the gauge
setting) is the prize to pay for maintaining off-shell positivity. Only
point-like (generally composite) local observables and the S-matrix are
independent of $e;$ but their perturbative computation requires the use of
string-local fields. The $p\cdot e\pm i\varepsilon $ terms in the
denominator are the momentum space expressions which correspond to the line
integrals in the creation respectively annihilation components.

The time-ordered propagators are formally obtained in terms of the
substitution%
\begin{equation}
\frac{d^{3}p}{2p_{0}}\rightarrow \frac{1}{2\pi }\frac{1}{p^{2}-m^{2}+i%
\varepsilon }d^{4}p
\end{equation}%
together with the Epstein-Glaser minimal scaling rule which allows the
appearance of undetermined counter-terms in case the scale dimension of the
propagator is $d\geq 4.$ For later use (section 3) we also note%
\begin{equation}
\partial ^{\mu }\int e^{-i\xi \cdot p}\frac{p_{\mu }p_{\nu }}{%
(p^{2}-m^{2}+i\varepsilon )(p\cdot \varepsilon +i\varepsilon )}=\partial
^{v}\int_{0}^{\infty }\delta (\xi +se)ds=:\partial ^{v}\delta _{e}(\xi )
\label{sdel}
\end{equation}

Whereas in the SLF Hilbert space formalism the particle creation and
annihilation operators are directly associated to Wigner-particle states,
the Krein space substitute$~$of the Proca potential%
\begin{equation}
A_{\mu }^{P,K}:=A_{\mu }^{K}-\partial _{\mu }\phi ^{K},~~\mathfrak{s}A_{\mu
}^{P,K}=0  \label{sub}
\end{equation}%
$~$yields only "emulates" such states \textit{inside matrix-elements} of
gauge-invariant operators. For the S-matrix one has

\begin{equation}
\left\langle q_{1}^{K}..q_{m}^{K}\left\vert S^{K}\right\vert
p_{1}^{K},..p_{n}^{K}\right\rangle =\left\langle q_{1}..q_{m}\left\vert
S\right\vert p_{1},..p_{n}\right\rangle  \label{scat}
\end{equation}%
where the Krein space vector meson states are obtained by successive
application of $A_{\mu }^{P.K}$ to the vacuum state$.$ The gauge formalism
can generally not prevent "leakage" of physical states defined by $\mathfrak{%
s}\left\vert \psi \right\rangle =0~$into unphysical regions of Krein space.

Besides the local observables, the only known shared global operator is the
S-matrix. A considerable conceptual difference is that in SLF in addition to
the perturbative definition of $S,$ there also exists a nonperturbative
derivation in terms of large time asymptotic properties of interpolating
fields in Hilbert space (on-shell unitarity from mass-shell restriction of
off-shell unitarity). In that case the $e$-independence of the S-matrix is a
result of the extension of scattering theory to space-like cone localized
operators (strings as limits of space-like cones) \cite{Haag}.

In both perturbative settings the calculations impose the requirement of $%
\mathfrak{s}$ or $d_{e}$ invariance on the formal time-ordered expression of
the Bogoliubov formalism. In the SLF setting one expects that the
perturbative S-matrix formalism can be extended to string-local physical
fields whose perturbative correlation functions are \textit{independent of
the }$e$\textit{-directions of internal propagators}. Such a distinction
between the string dependence of interacting fields, whose vacuum
expectation values one wants to calculate, and $e^{\prime }s$ which appear
in internal propagators of their perturbative expansion, has no counterpart
in gauge theory. Despite some formal analogies, the conceptual differences
between the global BRST cohomology of the $\mathfrak{s}$ and the geometric $%
d~$differential form calculus acting on the spacetime string directions
remain formidable.

\section{The breakdown of Haag duality and a new look at the Aharonov-Bohm
effect}

An example of an effect which cannot be described in the Krein space setting
of point-like vector potentials, but is correctly accounted for in the
string-local Hilbert space formalism, is the breakdown of Haag duality and
the closely related Aharonov-Bohm effect. It is instructive to illustrate
this in some detail in terms of Wilson loops.

For the following calculations it turns out to be convenient to work with
regularized electric and magnetic field strengths. Let $B$ be a small ball
centered at the origin and $\rho ,\sigma $ functions with supp$\rho \subset
B.$ Then we define regularized field strengths in terms of convolutions with 
$\rho ,\sigma $%
\begin{equation}
\vec{E}_{\rho }(\vec{x})=\int \vec{E}(\vec{x}-\vec{x}^{\prime })\rho (\vec{x}%
^{\prime })d^{3}\vec{x}^{\prime },~~\vec{H}_{\sigma }(\vec{x})=\vec{H}(\vec{x%
}-\vec{x}^{\prime })\sigma (\vec{x}^{\prime })d^{3}\vec{x}^{\prime }
\end{equation}%
The corresponding regularized fluxes through a surface $D$ are%
\begin{equation*}
E_{\rho }(D)=\int_{D}\vec{E}_{\rho }(\vec{x})d\vec{D},~\ H_{\sigma
}(D)=\int_{D}\vec{H}_{\sigma }(\vec{x})d\vec{D}
\end{equation*}%
Using the equal time commutation relation between the electric and magnetic
field strengths, one obtains for regularized electric and magnetic surface
fluxes through $D$ respectively $\hat{D}$%
\begin{eqnarray*}
&&4\pi i\left[ E_{\rho }(D),H_{\sigma }(\hat{D})\right] =\int \vec{g}%
_{D,\rho }(\vec{x})\func{curl}\vec{g}_{\hat{D},\sigma }(\vec{x})d^{3}\vec{x}
\\
&&\vec{g}_{D,\rho }(\vec{x})=\int \vec{g}_{D}(\vec{x}-\vec{x}^{\prime })\rho
(\vec{x}^{\prime }),~~\vec{g}_{D}(\vec{f})=\int_{D}\vec{f}d\vec{D}
\end{eqnarray*}%
where the vector-valued functions$~\vec{g}_{D,\rho }(\vec{x})$ and
correspondingly $\vec{g}_{\hat{D},\sigma }(\vec{x})$ are obtained by
regularizing the vector-valued surface distributions $\vec{g}_{D}(\vec{x})$
as define in the second line. Since the divergence of $\vec{E}_{\rho }(\vec{x%
})$ and$~\vec{H}_{\sigma }(\vec{x})$ vanishes, the corresponding flux $%
H_{\rho }(\hat{D})~$depends only on $\partial \hat{D};~$hence it can be
localized on any surface spanning $\partial \hat{D}.$

In a simple-minded picture of causality one expects that the fluxes are
localized on the tori~$\mathcal{T=\partial D+B},$ $\mathcal{\hat{T}=}%
~\partial \hat{D}+B,$ so that in case they interpenetrate but do not touch
the commutator vanishes.$~$But an explicit calculation for such a situation
shows that this is not true.

Taking for $D$ and $\hat{D}$ the discs 
\begin{eqnarray}
D &=&\left\{ \vec{x}\in \mathbb{R}^{3};~x_{3}=0,~x_{1}^{2}+x_{2}^{2}=1\right%
\} \\
\hat{D} &=&\left\{ \vec{x}\in \mathbb{R}^{3};~x_{1}=0,~\left( x_{2}-1\right)
^{2}+x_{3}^{2}=1\right\}  \notag
\end{eqnarray}%
whose associated interpenetrating tori $\mathcal{T=\partial }D+B,~\hat{T}=%
\mathcal{\partial }\hat{D}+B$ do not intersect for a sufficiently small $B,~$%
one finds%
\begin{equation}
4\pi i\left[ E_{\rho }(D),H_{\sigma }(\hat{D})\right] =\int \rho (\vec{x}%
)d^{3}x\cdot \int \sigma (\vec{x})d^{3}x  \label{RT}
\end{equation}%
This result for this straightforward but somewhat lengthy calculation has
been taken from old unpublished manuscript by Leyland, Robert and Testard 
\cite{LRT}. It is independent of the electric and magnetic surfaces $D~$and $%
\hat{D}$ as long as one does not change their boundaries.

The purpose of their calculation was to show the breakdown of Haag duality
for the system of localized operator algebras generated by a free field QED
field strength $F_{\mu \nu }(x).~$The terminology will be explained in the
sequel.

Causal localization properties of QFT are best described in the setting of
algebraic QFT in which models are defined in terms of their net of algebras$%
~ $of causally localized observables. Denoting an algebras localized in the
compact spacetime region $\mathcal{O}$ as $\mathcal{A(O)}$,$~$the causality
properties of QFT are expressed in terms of two independent relations 
\begin{eqnarray}
\mathcal{A(O}) &=&\mathcal{A(O}^{\prime \prime }),~~\mathcal{A(O})\subseteq 
\mathcal{A(O}^{\prime })^{\prime }  \label{Einstein} \\
\mathcal{A(O}) &=&\mathcal{A(O}^{\prime })^{\prime }\text{ }Haag~duality 
\notag
\end{eqnarray}%
Here the first equation is the causal completeness property ($\mathcal{O}%
^{\prime }$ is the causal complement and $\mathcal{(O}^{\prime })^{\prime }$
the causal completion); the second relation (in which the dash on the
algebra denotes its commutant) is the algebraic formulation of Einstein
causality; the special case of Einstein causality in the second line is
referred to as Haag duality \cite{Haag}.

The causal completeness property is a physically indispensable part of
causality, although it often does not receive the same attention as Einstein
causality\footnote{%
Isomorphisms between localized algebras in different spacetime dimensions
violate the causal completeness property, this affects in particular
Maldacena's AdS-CFT isomorphism \cite{Reh1} \cite{Reh2} \cite{SHPMP}..}. It
corresponds to the classical causal hyperbolic propagation.

Our intuitive idea about causality is through Haag duality; if there are
operators which violate Haag duality we tend to be surprised and seek an
explanation. The above observation about the existence of observables, which
can be localized on arbitrary (regularized) surfaces which share the same
boundary tori, can then be expressed as%
\begin{equation}
\vec{H}(\vec{\Phi}_{\hat{D},\rho })\subset \mathcal{A}^{\prime }(\mathcal{%
\hat{T}}^{\prime })\text{ }but\text{ }not~in~\mathcal{A(\hat{T}})
\end{equation}%
The magnetic flux commutes with all operators in $\mathcal{A(O}),~$with $%
\mathcal{O}$ a bounded contractible spacetime region which does not
intersect $\mathcal{\hat{T}}$ . One can always change the surface while
keeping its boundary in such a way that it is outside $\mathcal{O}$, but the
previous calculation shows that this is not possible if $\mathcal{O}~$is a
interpenetrating torus.

The tori and the regularized surfaces were constructed at a fixed time, but
using the causal completeness property, the relation (\ref{RT}) continues to
hold for their causal completion $\mathcal{T}^{\prime \prime },\mathcal{\hat{%
T}}$ $^{\prime \prime }~$in spacetime when the separation in space passes to
space-like separation in spacetime.

\textit{The use of string-local vector potentials in Hilbert space permits
an elegant explicit construction of such duality-violating operators.} For
this purpose one starts from the relation (\ref{rel}). In the massive case
the loop integral over the string-local vector potential is equal to the
that of its point-like counterpart. In the zero mass limit the Proca
potential and $\phi ~$do not exist; but since the differences $\phi
(x,e)-\phi (x,e^{\prime })$ remains infrared finite\footnote{%
I am indebted to Jens Mund for calling attention to this difference between
the massive case and its massless limit.}, the difference between two
identical Wilson loop but with different string directions vanishes%
\begin{equation*}
\doint A_{\mu }(x,e)dx^{\mu }-\doint A_{\mu }(x,e^{\prime })dx^{\mu }=0
\end{equation*}%
This leads to a kind of \textit{topological} $e$-dependence; the loop
integral still "remembers" that there was a directional dependence, but "it
forgets" the concrete space-like direction into which the $e$ pointed. This
"topological memory" of the Wilson loop corresponds to the breakdown of Haag
duality in QED.

One may picture this situation in terms of a semi-infinite \textit{cylinder}
formed by parallel space-like lines in the $e$ direction which emanate from
the points on the Wilson loop and extend to infinity. By $\rho $%
-regularization one can convert the circle into a torus, in which case the
wall of the cylinder has a finite thickness and the Wilson loop integral
represents a well defined operator to be used in localization arguments. The
semi-infinite cylinder is the covariant substitute of the surface and the
deformation of the latter corresponds to the Lorentz transformation of the
former. A magnetic loop which passes through the electric Wilson loop has to
penetrate the cylinder wall somewhere. An infinite extended magnetic flux
corresponds to a loop which closes at infinity.

This constructions of a regularized Wilson loop using a massless
string-local vector potential in Hilbert space leads to an elegant concrete
realization of an operator which is in $\mathcal{A(T}^{\prime })^{\prime }$
but not in $\mathcal{A(T)}$. On the other hand \textit{the gauge-invariant
Wilson loop defined in terms of the point-like vector potential of gauge
theory fails to account for the breakdown of Haag duality and the closely
related Aharonov-Bohm effect }(see below); the topological property is lost
in gauge theory. This was well-known to the authors of \cite{LRT} and for
this reason they avoided the use of point-local gauge potentials. In the
present formalism the vector potentials "live" in the same Hilbert space as
the field strengths.

Calculation in quantum mechanics are generally done in the Coulomb (or
radiation) gauge. It is interesting to note that this non-covariant and
non-local but \textit{rotation-invariant potential in Hilbert space} results
from the string-local potential by \textit{averaging over string directions} 
$e$ in a space-like hypersurface. Lacking covariance it plays no role in
covariant renormalization theory. The string-local vector potential is its
local covariant counterpart.

The quantum string-local vector potential has a classical analogue. It can
be obtained in terms of the expectation value of the quantum string-local
potential in coherent states. In this way the above operator calculations
pass to their classical counterpart; as argued in the sequel this shows that
the breakdown of Haag duality for free fields is basically a classical
phenomenon in which the localized subalgebras are replaced by modular
localized subspaces of the ($s=1,m=0)$ Wigner particle space \cite{BGL}.

In the classical setting the algebra $\mathcal{A}(\mathcal{T})$ corresponds
to $\mathcal{T}$-localized modular subspace $K(\mathcal{T})$ of the Wigner
space; the dash on the localization region retains its geometric meaning,
and the commutant of the algebra passes to the symplectic complement of the
subspace $K(\mathcal{T})~$(which is defined in terms of the imaginary part
of the inner product in Hilbert space \cite{BGL}). The classical analogue of
the breakdown of Haag duality is a topological phenomenon in classical field
theory. In this setting the $\mathcal{T}$ ~of the Aharonov-Bohm effect is
described in terms of a classical magnetic flux through an infinitely
extended solenoid which closes at conformal infinity. The role of quantum
mechanics in the A-B effect is the creation of the electric Wilson loop by
splitting an electron beam. Note that the topological semi-infinite cylinder
attached to the Wilson loop maintains a causal connection with the magnetic
flux whereas the point-local vector potential has no .

The breakdown of the Haag duality and its classical analog disappears in
case of massive vector mesons. Whereas the infrared properties are
characteristic for interacting mass-less vector-potentials, the duality
violation is a pure kinematic effect. Such topological duality violation
occur for all massless $s\geq 1$ tensor potentials.

The conformal invariance of the field strength raises the question whether
instead of space-like strings it would not be more appropriate to work with
light-like covariant string-local vector potentials; in this case the de
Sitter differential geometry would be replaces by that of the boundary of
the light-cone. Actually this could also lead to conceptual and
computational simplifications. Basic relations, as that between point-local
potentials and their string-local counterparts (the string-local potential
and its scalar escort $\phi $) seem to be preserved. We hope to return to
this interesting problem in a future publication.

Finally it is worthwhile to mention that the use of string-local potentials
also removes that "quirky feeling" about a missing causal relation between
the Wilson loop formed and the magnetic field passing through it. It
disappears once one realizes that it is caused by the use of point-like
vector potentials of gauge theory. This may be bad news for the popular
literature, but it is certainly helpful for demonstrating the power of
Hilbert space positivity (which, as shown before, even leaves its imprint in
the form of topological properties on the classical limit).

There exists another interesting scalar potential $\phi (x,e,e^{\prime
})\equiv \int_{0}^{\infty }e^{\prime \mu }A_{\mu }(x+\lambda e,e^{\prime
})d\lambda ^{\prime }$ which depends on two string directions and vanishes
on the diaginal $e=e^{\prime }$; hence it is localized on the space-like
wedge region spanned by the two space-like half-lines $\mathbb{R}_{+}e,~%
\mathbb{R}_{+}e^{\prime }.$ The important observation is that the zero mass
limit of suitable normalized exponential correlation functions of $\phi ~$%
remain finite for $m\rightarrow 0.$ This is reminiscent of the
two-dimensional situation which on encounters with conserved currents. The
curl of the related pseudo-current $\tilde{j}_{\mu }=\varepsilon _{\mu \nu
}j^{\nu }$~vanishes so that it permits a representation in terms of a
derivative of a scalar field~ $\tilde{j}_{{}}^{\mu }=\partial ^{\mu }\chi
(x) $. Again its zero mass limit diverges and the exponential functions
create new sectors in the massless limit.

The analogy may be described as follows%
\begin{eqnarray}
\chi (x,e) &=&\int_{0}^{\infty }\tilde{j}_{\mu }^{{}}(x+\lambda e)e^{\mu
}d\lambda \sim \phi (x,e,e^{\prime })=\int_{0}^{\infty }A_{\mu }(x+\lambda
e,e^{\prime })e^{\mu }d\lambda  \notag \\
&&\exp ig\chi (x,e)\sim \exp ig\phi (x,e,e^{\prime })  \label{infra}
\end{eqnarray}%
In both cases the zero mass limit of the exponentials generates new sectors
of the operator algebras generated by $j_{\mu }$ respectively $A_{\mu
}(x,e). $ The exponentials of $\chi $ play an important role in creating
"anyonic" sectors and "fermionization" in d=1+1. The exponentials functions
of $\phi $ on the other hand are expected to be important in a future theory
of "infraparticles"\footnote{%
The infrapaticle Dirac field is similar to the incoming free field, except
that it has the softerning of the electron mass shell already buildt in.}.
Another related shortcoming of the gauge theory in Krein space had been
noticed in \cite{Froe} namely that it leads to a vanishing Maxwell charge;
something which one does not expect in the Hilbert space setting. We hope to
return to this interesting problem.

\section{Perturbation theory in terms of string-local vector potentials}

More important for particle physics is the improvement of renormalizability
through the use of the string-local formalism. The simplest illustration of
this idea is provided by for massive QED. One rewrites the point-local
nonrenormalizable interaction density into its string-local counterpart as
follows%
\begin{eqnarray}
L^{P} &=&A_{\mu }^{P}j^{\mu }=(A_{\mu }-\partial _{\mu }\phi )j^{\mu
}=L-\partial ^{\mu }(j_{\mu }\phi )=L-\partial ^{\mu }V_{\mu }  \label{1} \\
&&\int L^{P}=\int L\longleftrightarrow ~L^{P}\overset{AE}{\simeq }L,\text{ ~}%
adiabatic\text{ }equiv.
\end{eqnarray}%
Here we have used the conservation of the free current (all fields are
non-interacting) and the notation $L$ for the string-local interaction
density. The power-counting violating point-like interactions has been
written in terms of a $L$ with $d_{sd}^{int}\leq 1$ and a $d_{sd}=5$
derivative $\partial V$ term which, at least in models with a mass-gap, can
be disposed of in the adiabatic limit of the first order S-matrix (second
line). The nontrivial step of generalizing this idea to higher order
time-ordered products will be undertaken in the next section.

Not all models are that simple. An interaction of a massive vector meson
with a Hermitian field $L^{P}=A^{P}\cdot A^{P}H$ leads to a $L,V~$pair $%
L-\partial ^{\mu }V_{\mu }=L^{P}$ (omitting the shared coupling strength $g$)

\begin{eqnarray}
&&L=m\left\{ A\cdot (AH+\phi \overleftrightarrow{\partial }H)-\frac{m_{H}^{2}%
}{2}\phi ^{2}H\right\} ,~V_{\mu }=m\left\{ A_{\mu }\phi H+\frac{1}{2}\phi
^{2}\overleftrightarrow{\partial }_{\mu }H\right\}  \notag \\
&&Q_{\mu }=d_{e}V_{\mu }=mu(A_{\mu }H+\phi \overleftrightarrow{\partial }%
_{\mu }H)  \label{2}
\end{eqnarray}%
In this case the string-local interaction density (and not only the the $%
V_{\mu }$ as in massive QED)~depends on $\phi .~$There are other terms
within the power-counting restriction which we could have added namely $%
cH^{3}+dH^{4}$ with initially independent coupling strengths $c,d$. But it
turns that the second and third order $e$-independence of the S-matrix 
\textit{induces} these couplings anyhow, as well as additional $H$-$\phi $
couplings. The basic difference of this kind of induction as compared to the
standard counterterm formalism is that it does not increase the number of
parameters. In fact the first and second order induced $H$-selfcouplings and 
$H$-$\phi $ couplings taken together turn out to have the form of a Mexican
hat potential. Its appearance has nothing to do with SSB; it is fully
explained in terms of the $e$-independence of the S-matrix (next section) .

Here the terminology \textit{induced} refers to contributions whose presence
is required for the $e$-independence of scattering amplitudes. This is a
consequence of scattering theory in the presence of a mass gap; we refer
again to a general theorem in \cite{Bu-Fr2} \cite{Haag} which states that
the LSZ scattering theory permits an extension to string-local fields and
that the difference between point- and string-localized fields disappears on
the level of incoming/outgoing particle states. In the context of the gauge
theoretical formalism in Krein space there are no physically localized
fields in Hilbert space to which this theorem can be applied; hence the
imposition of gauge invariance on the perturbative representation of the
scattering amplitude remains a perturbative prescription.

The Krein space of gauge theory does not contain multiparticle Wigner
states; the best one can do is to view the free fields $A^{P,K}$ as the 
\textit{substitute} for the generating fields $A^{P}$ of the Wigner
particles in the sense of (\ref{scat})\footnote{%
The speculative remarks in \cite{Du-S} on the gauge theoretic substitutes of
Wigner particles acquire a concrete meaning in the present setting.}. The
field-particle connection resulting from scattering theory and problems
concerning the relation of the elementary (model-defining) fields with the
(generally composite) local observables and their associated bound states
remain outside its physical range.

A useful reformulation for the construction of $e$-independent first order
interaction densities in terms of a given number of string-local fields is
to look directly for $L,V_{\mu }$ or $L,Q_{\mu }~$pairs with 
\begin{equation}
d_{e}(L-\partial ^{\mu }V_{\mu })=0~or~d_{e}L-\partial ^{\mu }Q_{\mu
}=0,~~d_{sd}^{L}\leq 4  \label{V}
\end{equation}%
instead of starting from an \thinspace $L^{P}.$ The $L~$of such a pair is
then determined up to exact forms $d_{e}C$ and the $V_{\mu }$ up to a
divergence-free current.~The definition of point-like interaction densities
in terms of such pairs is particularly useful for the generalization to
higher order time-ordered point-local interaction densities.

Whereas for theories with a mass gap the construction of a $L,V_{\mu }~$pair
and the extension to $e$-independent time-ordered higher order densities has
a clear motivation, this is lost in the massless limit when $\phi $ and$%
~V_{\mu \text{ }}$ and the S-matrix cease to exist (become infrared
divergent). This, as well as the appearance of unexpected new phenomena, as
those presented in the previous section, leaves only one way to construct
massless $s=1$ interacting fields. It consists in taking the $m\rightarrow 0$
limit of the correlation functions of the associated massive theory and
construct the massless operator theory in Hilbert space from its correlation
functions.

In this way one avoids the intermediate use of infrared divergent operators
as $V_{\mu },Q_{\mu }$ and one also bypasses the problem of an explicit
description the zero mass Hilbert space whose structure is very different
from that of a Wigner-Fock Hilbert space.~Wightman's reconstruction theorem
insures that a Hilbert space and quantum field operators can be constructed
from (positivity-obeying) correlation functions of interacting \cite{St-Wi}.

The use of string-local charge-carrying matter fields is supported by
rigorous results in QED \cite{B}. Whereas in the presence of a mass gap
point-local matter fields may still exist in the form of singular fields
with unbounded short distance dimensions and therefore outside the setting
of localizable Wightman fields, the massless limit only contains
string-local fields and point-local observables.

The BRST gauge formalism uses charge-carrying unphysical point-like fields.
This limits its physical range to local observables and the the perturbative
S-matrix\footnote{%
Tthe physical on-shell scattering operator S agrees with Bogoiubov's
generating off-shell S functional only in the adiabatic limit.}. Its
perturbative construction has been presented in \cite{Scharf}. It can be
formulated in a setting which parallels the of $L,Q$ pairs%
\begin{eqnarray}
&&\mathfrak{s}L^{K}-\partial ^{\mu }Q_{\mu }^{K}=0 \\
S &=&\int L^{K}(x)d^{4}x,~~\mathfrak{s}S=0
\end{eqnarray}%
where $K$ refers to Krein space. Integration over Minkowski spacetime
removes the $Q$ contribution in the adiabatic limit so that the S-matrix
only depends on $L^{K}$. The formal expressions in the case of coupling to a
Hermitian matter field $H$ parallel those of (\ref{2}) with the fields
replaced by their Krein space counterparts.

The BRST formalism, apart from the vacuum sector generated by gauge
invariant observables, implements only on-shell unitarity. Gauge theory can
describe scattering, but it does not permit to extend on-shell unitarity to
off-shell positivity; properties abstracted from gauge-dependent fields have
no physical content. In most applications of QFT to particle physics the
S-matrix in the presence of a mass gap covers all problems of interests. But
in massless $s=1~$theories as QED, for which the S-matrix ceases to exist,
one is forced to use additional momentum-space recipes which repair the
infrared divergencies of the perturbative scattering amplitudes and obtain
instead finite photon-inclusive cross sections.

A spacetime understanding of "infraparticles" in QED does not yet exist, not
to mention the much harder problem of large distance behavior leading to
confinement of gluons and quarks in QCD. Such problems are outside the range
of gauge theory. They require the understanding of large distance properties
of physical fields. For this one needs to extend the unitarity of the
S-matrix ("on-shell unitarity") to the Hilbert space description ("off-shell
unitarity") of fields. This cannot be achieved within the setting of
point-local fields since point-like localization for interactions involving
vector mesons is incompatible with the Hilbert space positivity.

As described before, the way out is to use the structural simplicity of a
Wigner-Fock Hilbert space description in the presence of a mass-gap and to
define the massless theory in terms of the $m\rightarrow 0~$limit of the
correlation functions. In this way one maintains the positivity in the
massless limit so that one can study the long distance behavior in terms of
physical string-local fields.

A particularly interesting case arises if confinement in QCD results from
the vanishing of those correlation functions which formally contain besides
point-like (hadronic and gluonium) composites and "string-bridged" $q-\bar{q}
$ pairs also string-local gluons and quarks. In that case one obtains a QFT
in which the basic fields (those used in the definition of the perturbative
first order interaction density), have disappeared and only their
point-local composites (gluonium, hadrons) remain. The only way to describe
such a situation within our present understanding of QFT is to view it as a
massless limit of a massive theory in which the particles of the model
defining elementary fields appear in the Wigner-Fock Hilbert space (the last
section returns to this problem).

An analogue of (\ref{rel}) for arbitrary spin exists for all $s\geq 1;$
instead of string-local scalar $\phi $ one obtains a linear relation
involving derivatives of~string-local escort fields $\phi $ of spin $<~s$
with short distance dimension $d_{sd}<$ $s+1$. It is not clear whether this
can be used to generalize the idea of obtaining short distance improving $%
L,\partial V$ pairs to higher spin. A particularly interesting case is $s=2.$

\section{Induced higher order contributions and the Higgs model}

The construction of a string-local renormalizable first order interaction
density in terms of a $L,V_{\mu }$ pair permits an extension to higher
orders. The second order relation ($L^{\prime }$ stands for $L(x^{\prime
},e^{\prime })$)%
\begin{equation}
(d_{e}+d_{e^{\prime }})(TLL^{\prime }-\partial ^{\mu }TV_{\mu }L^{\prime
}-\partial ^{\prime \nu }TLV_{\nu }^{\prime }+\partial ^{\mu }\partial
^{\prime \nu }TV_{\mu }V_{\nu }^{\prime })=0  \label{indep}
\end{equation}%
would automatically follow from the first order relation (\ref{1}) if it
would not be for the singularities coming from time-ordering of distributions%
\footnote{%
A systematic presentation of the differential form calculus will be
contained in forthcoming work by Mund.}. If this renormalization condition
can be implemented, the first order definition (\ref{V}) of the point-like
interaction density permits a second order generalization 
\begin{equation}
TL^{P}L^{P\prime }:=(TLL^{\prime }-\partial ^{\mu }TV_{\mu }L^{\prime
}-\partial ^{\prime \nu }TLV_{\nu }^{\prime }+\partial ^{\mu }\partial
^{\prime \nu }TV_{\mu }V_{\nu }^{\prime })  \label{p}
\end{equation}%
and one could hope to be able to generalize this idea to higher orders.

It is instructive to recall how the CGI gauge theoretic procedure \cite%
{Scharf} deals with this problem. The formal proximity to the string-local
setting is evidenced in case one uses the weaker "$Q$-version"%
\begin{equation}
\mathfrak{s}TL^{K}L^{K\prime }-\partial ^{\mu }TQ_{\mu }^{K}L^{K\prime
}-\partial ^{\prime \mu }TL^{K}Q_{\mu }^{K\prime }=0  \label{K}
\end{equation}%
Replacing the BRS$~\mathfrak{s}~$by $d=d_{e}+d_{e^{\prime }}$ and omitting
the superscript $K,$ one obtains the corresponding SLF relation. Although
this $Q$-relation is weaker than the $V$-relation, it suffices to define a
second order gauge-invariant ($\mathfrak{s}S=0)$ or string-independent ($%
dS=0 $) S-matrix since the derivative terms drop out in the adiabatic limit.

Starting from time ordering $T_{0}$ in which all derivatives of fields are
taken outside, one can use the Epstein-Glaser minimal scaling prescription
to define a renormalized parameter-dependent $T$-products. For a scalar free
field of scale dimension $d_{sd}=1$ the $T$-ordering for the $d_{sd}=2~$%
derivative is%
\begin{eqnarray}
&&\left\langle T_{0}\partial _{\mu }\varphi (x)\partial _{\nu }^{\prime
}\varphi ^{\ast }(x^{\prime })\right\rangle =\partial _{\mu }\partial _{\nu
}^{\prime }\left\langle T_{0}\varphi (x)\varphi ^{\ast }(x^{\prime
})\right\rangle  \notag \\
&&\left\langle T\partial _{\mu }\varphi (x)\partial _{\nu }^{\prime }\varphi
^{\ast }(x^{\prime })\right\rangle =\left\langle T_{0}\partial _{\mu
}\varphi (x)\partial _{\nu }^{\prime }\varphi ^{\ast }(x^{\prime
})\right\rangle +cg_{\mu \nu }\delta (x-x^{\prime })  \label{T}
\end{eqnarray}%
where the second line is the definition of a one-parametric $T$-ordering
according to the E-G minimal scaling rule. It turns out that this freedom
can be used to absorb certain "anomaly" contributions into a redefinition of
time-ordering. The strategy is to use the freedom in the definition of $T$
in such a way that (\ref{K}) is fulfilled.

One defines an anomaly as a measure of violation of (\ref{indep}) or (\ref{K}%
) if one uses $T_{0}$ instead of the still unknown $T.$

\begin{eqnarray}
\mathfrak{A}_{V} &:&=(d_{e}+d_{e^{\prime }})(T_{0}LL^{\prime }-\partial
^{\mu }T_{0}V_{\mu }L^{\prime }-\partial ^{\prime \nu }T_{0}LV_{\nu
}^{\prime }+\partial ^{\mu }\partial ^{\prime \nu }T_{0}V_{\mu }V_{\nu
}^{\prime })  \label{A1} \\
\mathfrak{A}_{Q} &:&=(d_{e}+d_{e^{\prime }})T_{0}LL^{\prime }-\partial ^{\mu
}T_{0}Q_{\mu }L^{\prime }-\partial ^{\prime \mu }T_{0}LQ_{\mu }^{\prime }=0
\label{A2}
\end{eqnarray}%
For the calculation of two-particle scattering we only need the 1-particle
contraction component (the "tree" approximation).

The simplest nontrivial illustration is provided by massive scalar QED%
\footnote{%
The renormalization theory of massive spinor QED has no anomalies.}. In that
case the presence of a derivative in the current (\ref{1}) leads to delta
function contributions from the divergence of the two-point function of the
charged field $\varphi $%
\begin{equation}
\partial ^{\mu }\left\langle T_{0}\partial _{\mu }\varphi (x)(\partial _{\nu
}^{\prime })\varphi ^{\ast }(x^{\prime })\right\rangle =\partial _{\nu
}\delta (x-x^{\prime })+reg.  \label{A3}
\end{equation}%
where the regular part come from the use of the free field equation inside
the $T_{0}.~$This together with $d_{e}(\partial \phi \cdot A)=\frac{1}{2}%
d_{e}(A\cdot A),$ which results from the application of $d_{e}~$to (\ref{rel}%
), yields \cite{FOOP}

\begin{eqnarray}
&&\mathfrak{A}_{Q}=d_{e}T_{0}LL^{\prime }\mathfrak{-}d_{e}N_{sym}+\partial
_{\mu }N_{sym}^{\mu },~N_{sym}=N+N^{\prime },~N_{sym}^{\mu }=N^{\mu }+N^{\mu
\prime }  \label{a} \\
&&N=\delta (x-x^{\prime })\varphi ^{\ast }\varphi A\cdot A^{\prime },~N_{\mu
}=\delta (x-x^{\prime })\varphi ^{\ast }\varphi \phi A_{\mu }^{\prime }
\end{eqnarray}%
The $N^{\prime }$ and $N_{\mu }^{\prime }~$are obtained by symmetrization $%
x,e\longleftrightarrow x^{\prime },e^{\prime }.~$Using the freedom (\ref{T}%
), the $N$ and $N_{\mu }~$can be absorbed into a redefinition of of
time-ordering 
\begin{eqnarray}
&&TLL^{\prime }=T_{0}LL^{\prime }+N_{sym},~TQ_{\mu }L^{\prime }+TLQ_{\mu
}^{\prime }=T_{0}Q_{\mu }L^{\prime }+T_{0}LQ_{\mu }^{\prime }+N_{\mu ,sym}~
\label{c} \\
~ &&S^{(2)}=-\frac{g^{2}}{2}\int \int TLL^{\prime }d^{4}xd^{4}x^{\prime }~
\end{eqnarray}%
In this way the SLF counterpart $dS=0$ of (\ref{K}) has been established.
The derivation of the second order point-like density (\ref{p}) is more
involved but it does not change $TLL^{\prime }.$

The result is hardly surprising since the presence of a quadratic term in
the vector potential from $N_{sym}~$in second order is well known from gauge
theory. But there is a subtle point of fundament significance which should
be mentioned. In general it is not possible to set $e=e^{\prime }$ The
reason is that string-local fields $\Psi (x,e)~$\ fluctuate in both $x$ and $%
e$ and hence products of fields with the same $e$ do not make sense inasmuch
as coalescing values of $x$ in products of fields diverge. Fortunately
Wick-ordered products do not only permit the $x$ to coincide, but they also
allow $e^{\prime }s$ to coalesce$.~$Such $e^{\prime }s$ inside Wick products
will be called "mute". But the expansion of time-orderd products into
Wick-products also involves inner time-ordered contractions for which the $e$%
-fluctuations prevent $e^{\prime }s$ from coinciding.

Consider the propagator of a string-local vector meson.

\begin{equation}
\frac{1}{p^{2}-m^{2}}(-g_{\mu \nu }+\frac{p_{\mu }p_{\nu }}{(p\cdot
e-i\varepsilon )(p\cdot e^{\prime }+i\varepsilon )}+...)  \label{prop}
\end{equation}%
As a result of the different $i\varepsilon $-prescriptions, the
distributional boundary values are ill-defined for $e=e^{\prime }$. This is
a problem which cannot be solved by renormalization theory. It explains why
the axial gauge (which treats the $e$ as a rigid gauge parameter) had to
fail.

In the present context the $e^{\prime }s~$are fluctuation spacetime
variables of a string-local covariant field. The implementation of the
independence of the S-matrix of the string directions $dS=0$ based on the
differential calculus of the $1+2~$dimensional de Sitter space guaranties
that this problem disappears after adding sufficiently many on-shell
contributions. The simplest illustration is that of second order scattering
of charged particles. In that case the dangerous contribution comes from the
vector meson propagator (\ref{prop}). But the contribution containing both $%
e^{\prime }s$ disappears after the use of the on-shell current conservation%
\footnote{%
The argument parallels that of second order gauge-invariance in gauge
theories.} and the remaining $e$-dependence is then canceled by the
contribution from $N_{sym}.$

It is not necessary to go into the details of these cancellations since the
relation $dS=0,~$which follows in the adiabatic limit from (\ref{c}),
guaranties $e$-independence. But as individual perturbative contributions to
gauge-invariant scattering amplitudes in gauge theories are generally not
gauge invariant, it is hardly surprising that individual contributions in
the SLF setting still depend on $e.$ What is however somewhat unexpected is
that such contributions may diverge for coalescent $e^{\prime }s$. The
differential calculus in the $1+2$ dimensional de Sitter space of string
directions only guaranties that the sum of perturbative contributions in a
particular order converges. A formal off-shell continuation by hand would
destroy this result; only the extension of perturbation theory to fields can
preserve the independence from $e$-fluctuations resulting from inner
propagators.

This observation contains an important message. The positivity of Hilbert
space, which required the use of string-local fields in order to preserve
renormalizability, can be implemented in such a way that the scattering
matrix remains string-independent. But in contrast to gauge theory
individual contributions to $S$ may depend on the $e^{\prime }s$ in such a
way that equating different $e^{\prime }s$ causes infinite fluctuations in
separate terms.

On the basis of these observations one anticipates that an extension of this
formalism to interacting string-local fields will lead to perturbative
contributions which depend on the individual string-directions of inner
propagators after all $p$-integration have been carried out. But one expects
that in suitable sums over contributions the dependence on the inner $%
e^{\prime }s$ disappears so that only the possible string dependence of the
interacting fields remains. This is very different from gauge theory; the
BRST $\mathfrak{s}$ is a global operation which cannot distinguish between
internal and external propagators

The Hilbert space setting, which requires the use of the differential
calculus in the directional de Sitter space, is certainly more elaborate
than the gauge formalism, but unlike the latter it is not limited to the
construction of local observables and the S-matrix but it also leads to the
construction of all string-local fields.

After this interlude of conceptual difference of the SLF setting with
operator gauge theories presented in the contest of second order scalar
massive QED, we now turn to the problem of the second order calculation of
the $A$-$H~$coupling (\ref{2}). Up to $H$ self-interactions the $L,V~$pair
is uniquely determined by first order power-counting $d_{sd}^{int}\leq 4$
and the requirement $d_{e}(L-\partial V)=0.$

Our aim is to highlight differences between the gauge theoretic second order
calculations as it was presented in \cite{Scharf} \cite{Aste} and the SLF
setting. In that work the calculation of the second order gauge invariant
for the Higgs model S-matrix starts with a first order $L^{K},Q_{\mu }^{K}$
pair which is identical to (\ref{2}), except that $L,~A_{\mu },$ $\phi ~$and 
$Q_{\mu }~$now have the superscript $K$ and instead of the geometric
differential$~$calculus one now has the abstract $\mathfrak{s}$-operation.
The renormalizable $cH^{3}+dH^{4~}$terms which were already mentioned after (%
\ref{2}) turn out to be necessary in order to keep the third order tree
contributions free of anomalies; beyond the third order there are no tree
anomalies.

The result is \cite{Scharf}

\begin{eqnarray}
&&T_{0}L^{K}L^{K\prime }+i\delta (x-x^{\prime })(A^{K}\cdot
A^{K}H^{2}+A^{K}\cdot A^{K}\phi ^{2})-i\delta (x-x^{\prime })R^{K}
\label{second} \\
&&R^{K}=-\frac{m_{H}^{2}}{4m^{2}}(\phi ^{K,2}+H^{2})^{2},~V^{K}=g^{2}\frac{%
m_{H}^{2}}{8m^{2}}(H^{2}+\phi ^{K,2}+\frac{2m}{g}\phi ^{K})^{2}-\frac{%
m_{H}^{2}}{2}H^{2}  \notag
\end{eqnarray}%
where $V^{K~}$is the contribution to the second order S-matrix which results
from combining the second order $R^{K}$ with the first order $H$%
-selfinteractions. The appearance of a $1/g$ term shows that this way of
writing is somewhat artificial, but it permits to relate the physical
parameters $m,m_{H},g$ to those which result from applying a shift in field
space (the abelian SSB Higgs-mechanism) to the gauge dependent scalar field
of QED.

All interactions which involve point-like vector mesons require the setting
of gauge theory, including those in which the vector meson interacts with
Hermitian fields. The result (\ref{second}) shows that there is no place for
SSB. Whereas this is possible for selfinteracting scalar fields, it is ruled
out when scalar fields also couple to vector mesons. In that case the vector
mesons require the implementation of the rules of gauge theory, which leaves
no room for SSB. The coupling to string-local vector potentials in the SLF
Hilbert space formulation converts the point-local scalar fields into
interacting string-local fields whereas SSB is limited to point-local scalar
fields.

This is confirmed by the second order ghost-free SLF Hilbert space setting
which only uses the causal localization principles of QFT. Instead the
abstract cohomological BRST formalism it is based on the geometric
differential form calculus in the de Sitter space applied to string-local
fields with fluctuating space-like directions. The calculation of a $e$%
-independent S-matrix follows similar steps; the details will be presented
elsewhere.

There is however one difference between the gauge theoretic calculation and
its SLF analogue which is worthwhile mentioning. In addition to the induced
potential $R$ there is a second order term of the form

\begin{equation}
\delta (x-x^{\prime })A_{\mu }^{P}A^{\mu \prime }u(\phi ^{\prime }-\phi
)+(e\longleftrightarrow e^{\prime })  \label{special}
\end{equation}%
The short distance dimension of the point-local Proca field is $%
d_{sd}^{P}=2,~$this contribution has $d_{sd}=5~$and hence violates the power
counting bound. But since it vanishes for $e=e^{\prime }$ it causes no
problem for the $e$-independent S-matrix.

The Schwinger-Swieca screening effect which asserts that any interaction
which involves massive vectormesons (independent of whether it couples to
charged or Hermitian matter) leads to a vanishing Maxwell charge, is a
rigorous argument against any form of SSB interpretation of the Higgs model.
It generalizes the screening of the Maxwell current of a free massive field
strength

\begin{equation}
j_{\nu }:=\partial ^{\mu }F_{\mu \nu }=m^{2}A_{\nu }^{P},~~Q_{Max}=\int
A_{0}^{P}(x)d^{3}x=0  \label{max}
\end{equation}%
which can easily be checked by looking at the form of the two-point function
of the free Proca field. Higher order contributions to $F_{\mu \nu }~$%
require an extension of the $S$-matrix formalism to interacting fields, but
thanks to the Swieca theorem, this is not needed if one only wants to
exclude a SSB mechanism.

If the Higgs model would be the result from a SSB, the charge would diverge $%
Q=\infty $ instead of vanishing. This is part of the definition of SSB. A
symmetry is called "spontaneously broken" if the Noether theorem cannot be
inverted i.e. when a conserved current $\partial ^{\mu }j_{\mu }=0$ leads to
a a divergent charge (the generator of a symmetry), formally $Q=\int
j_{0}(x)d^{3}x=\infty .$

A SSB model is closely related to a symmetric theory, in fact both are
unitarily inequivalent representations ("different vacua") but their local
algebras remain equivalent \cite{E-S}\thinspace . In $s=1~$indefinite metric
gauge theories these concepts are not available and in the Hilbert space
formulation the string-local coupling of a massive vector meson to a
Hermitian scalar field bears no local equivalence to one with a complex
matter field. The screening phenomenon of massive vectormesons is
independent of the matter to which it couples. As soon as scalar matter
couples to massive vector mesons the rules of the SSB game change in that
the massive vectormesons prevent the SSB of scalar self-coupled particles
and enforces the screening property regardless of whether the scalar matter
is complex (scalar QED or the Higgs model).

Unlike for self-interacting scalar particles without vectormesons where
there exist two "phases" (symmetric and SSB) there is simply no other
interaction of scalar fields than massive scalar QED and its Hermitian Higgs
counterpart without any physical SSB bridge between them. This does not
exclude the possibility of a calculational recipe which relates a massless
vector potential with a complex scalar to the Higgs model. Any manipulation
which leads to a massive vector meson and a Hermitian field will end up as
the Higgs model since any renormalizable model with all its first order
interactions terms is uniquely dermined by its field content. If one forgets
a term contributing to the interaction density then second order "induction"
(by gauge invariance or $e$-independence) and higher order counterterms from
renormalization will lead to that unique theory with that field content. The
physical nature of that theory (SSB or screening) can only be unravelled in
terms of intrinsic properties (as the preserved local symmetry in case of
SSB).

The most interesting situation is that of \textit{massive selfinteracting
vector mesons}. A first order $L^{K},Q_{\mu }^{K}~$pair within the
power-counting limit is easily found, but it is not possible to maintain
this renormalizability restriction in the calculation of the second order
gauge-invariant S-matrix. As already mentioned in the introduction, one can
only compensate the "bad" terms by introducing a first order coupling with
scalar $H$-fields which in second order produces a compensating bad terms.
The calculations have been carried out in the CGI setting of gauge theory in 
\cite{Scharf} \cite{Aste}; these authors also showed that in case of just
one $H$ the model contains no additional parameters besides one coupling
strength and the masses of $H~$and the vectormesons. In \cite{BDSV} this
situation was reviewed and compared with the actual situation of the
Standard Model.

This result is confirmed in the SLF Hilbert space setting. The details
require more extensive calculations and will be presented in a joint paper
with Jens Mund. There is one difference with the CGI calculation which is
worth mentioning. The calculation of the second order S-matrix from the
first order $A$-$H$ coupling 
\begin{equation}
L_{H}=\sum_{a,b}d_{a,b}(A_{\mu ,a}^{P}A_{b}^{\mu }H+A_{a}^{\mu }\phi
_{b}\partial _{\mu }^{{}}H-\frac{1}{2}m_{H}^{2}\phi _{a}\phi _{b})
\end{equation}%
parallels that of the abelian coupling, but instead of the term (\ref%
{special}) which has short distance dimension $d_{sd}=5$ instead of $4$ (but
fortunately vanishes on the $e$-diagonal $e^{\prime }=e)~$one now finds%
\begin{equation}
\delta (x-x^{\prime })\sum_{a,b;a^{\prime }b^{\prime }}d_{a,b}d_{a,^{\prime
}b^{\prime }}u_{a}A_{b^{\prime }}^{\mu }(A_{\mu ,b}^{P}\phi _{a^{\prime
}}^{\prime }-A_{\mu ,a^{\prime }}^{P}\phi _{b})+(e\longleftrightarrow
e^{\prime })  \label{special2}
\end{equation}%
which does not vanish on the $e$-diagonal. Hence the second order
contribution to the $e$-independent S-matrix contains a
power-counting-violating term which can only be compensated by a similar
term from the second order selfinteraction between the massive vector
mesons. The full calculation in the new Hilbert space setting will be
contained in forthcoming work.

As long as one considers the shift in field space and the subsequent gauge
transformation as formal devices which relate the well-known Lagrangian of
scalar QED with a new Lagrangian in which a massive vector meson interacts
with a H-field, no harm is being done. The problem only starts if one
attributes a physical interpretation with this recipe. In this case one
misses a new physical mechanism namely the loss of second order
renormalizability of selfinteracting massive vector mesons which can only be
saved by extending the first order interaction by an additional coupling
with a$~H$-field which produces second order compensating terms.

Such a compensatory mechanism between fields with different spin which
preserves renormalizability was hoped for to occur in models with
supersymmetrc couplings. Whereas such a requirements played no role in the
definition of supersymmetry, it is the raison d'\^{e}tre for the $H$%
-particle. Formal manipulations of Lagrangians contain generally no physical
information; the physical content of a model in QFT is always related to
intrinsic properties of the model independent of the way in which it was
constructed.

Terminologies as "fattening" or "being eaten" should have served as a
warning that one is entering metaphoric swampland and reminded particle
theorists that understanding of properties of QFT means connecting them with
the foundational causal localization principle of QFT. We leave it to the
historians to explain why, despite the correct terminology "Schwinger-Higgs
screening" in some publications at the time of the Higgs paper the related
ideas where lost in the maelstrom of time.

One of the reasons which contributed to its popularity may be related to the
fact that the Higgs mechanism was discovered in at least three independent
papers with identical metaphors about the Goldstone boson "being eaten by
the massless vector meson". The problem here is not that important
discoveries have been made by metaphoric ideas; this happened many times in
the history of physics. What is however a bit embarrassing is that even 40
years later these reasonings appear in important documents.

For the sustentation of the impressive experimental effort, which after
decades of search finally led to the discovery of $H~$at LHC, the narrative
about particles, which in addition of generating masses of vector mesons
also create their own mass, may have been a blessing; it would have been
much harder to convince experimentalists that the fate of the Standard Model
depends on the need to find an additional particle in order to uphold second
order renormalization. The fundamental aspect of such an observation is more
concealed since it is related to the not yet sufficient understood
connection between renormalizability with the causal localization principle
in a Hilbert space setting.

\section{Spacetime oigin of the Lie structure of selfinteracting vector
mesons}

One of the unsolved problems of $s=1$ interaction is the understanding of
the conceptual origin of the quantum Lie structure in models of
selfinteracting vector mesons. The answer cannot be given by referring to
gauge theory. In classical gauge theory the Lie algebra structure is part of
the fibre bundle formalism whereas the only principle of QFT is causal
localizability. The BRST gauge setting is a compromise between the ideas
underlying classical gauge invariance and the quantum requirements. A
derivation of the quantum Lie structure within the BRST gauge setting, as
can be found in \cite{Scharf}, is hardly surprising. It should rather be
understood \textit{as a consequence of the foundational causal localization
properties for selfinteracting }$s=1$\textit{\ fields.}

Quantum gauge-symmetry cannot be realized in a Hilbert space; the
preservation of the classical gauge structure requires the use of an
indefinite metric Krein space and hence the gauge-theoretic derivation of
the Lie algebraic structure in a BRST gauge setting involves the use of a
circular argument.

The understanding of the physical concepts behind \textit{local} gauge
symmetry turned out to be one of the hardest problems of local quantum
physics. The origin of \textit{global} gauge symmetries (inner symmetries)
had already been well understood during the 70s in terms of the DHR and the
more general later DR superselection theory \cite{Haag}. These constructions
show that observable algebras, which typically arise as invariant
subalgebras of field algebras under the action of a global symmetry group,
contain sufficient information for reconstructing the symmetry group and the
field algebra on which it acts.

This construction is somewhat astonishing since at first glance the causal
localization properties of local observables seem to have no connection with
group representation theory; the net of localized observable algebras is
covariant under the Poincar\'{e} group (including the TCP operation) but it
is no relation to gauge symmetry, be it global or local. Yet the
classification of unitarily inequivalent local representations of the
spacetime-indexed set of local algebras (the local superselection sectors of
the observable algebra) leads indeed to a field algebra and a compact groups
acting on it. The construction uses only the spacetime causal localization
properties; the group theory is hidden in the composition structure ("fusion
rules") of the localization-preserving inequivalent representations
(endomorphisms) of the observable algebra \cite{Haag}. In this way the
origin of global gauge symmetries (inner symmetries) is fully accounted for.

These methods work in the presence of a mass gap but fail for interacting
theories involving massless vector potentials. \textit{Local gauge symmetry
tries to imitate global gauge symmetry at the prize of loosing the Hilbert
space}. There have been attempts to classify representation sectors of the
algebra of local observables of QED \cite{Bu-Ro}. They require the
introduction of new concepts and their results are presently incomplete and
far from being useful for the understanding of the problems of interacting
massless vector mesons.

An understanding of the Lie group structure of selfinteractiong vector
mesons would be a useful step in this direction. Selfcouplings between $s<1$
fields are not subject to such restrictions. Forming renormalizable
trilinear and quadrilinear couplings between a finite set of low spin fields
allows a very large number of independent coupling parameters. For any
symmetric low spin $s<1$ model with one coupling parameter there exists a
large number of less symmetric models with many independent couplings
between the same free fields. So why is this not possible for $s=1~?.$ The
answer is that \textit{for }$s\geq 1$\textit{\ the Hilbert space positivity
together with renormalizability is more restrictive.}

Instead of an interaction density $L$ one has to find an $e$-dependent $%
L,V_{\mu }~$pair which satisfies a differential relation in $e$. The local
counterterms of the inductive Epstein-Glaser formalism, which for $s<1$
interactions between point-local fields are only subject to the the
appearance of point-like counterterms at coalescent $x$ restricted by the
minimal scaling requirement, have now to fulfill additional differential
relations from string-crossings. These additional restrictions originate
from maintaining causality in the presence of string-local field; they are
not imposed symmetries. \textit{This more restrictive setting accounts for
the appearance of Lie-algebraic structures in interactions involving }$s=1.$

The formalism of local gauge theory on the other hand results from trying to 
\textit{emulate} this new structure \textit{within a point-local setting}.
The prize is the loss of the Hilbert positivity and the "gain" is the
promotion of the Lie-algebra structure of the coupling of fields to a full
"local gauge symmetry". In this way the formalism is formally reunited with
that for $s<1$ interactions at the prize of its smaller physical range.
Important problems of QED and QCD, which require the understanding of
long-distance behavior of fields, remain outside the physical range of local
gauge theory.

These new ideas are presently limited to perturbation theory. But they
suggest that the nonperturbative construction of appropriately defined
string-local field algebras, which are uniquely associated with the local
observable algebras of massless $s=1$ interactions, may be a realistic goal
of a project as that in \cite{Bu-Ro}. This requires to extend the
point-local Wightman setting to string-local fields $\Psi (x.e)~$which are
tempered distributions in both $x$ and $e.$

As in previous second order calculations one expects that the formal
similarity with the gauge formalism in \cite{Scharf} extends to the model of
selfinteracting vector mesons. The following calculations shows that this is
indeed the case.

For simplicity we restrict our calculation to the massless first order
interaction. The starting point is the renormalizable $L,Q_{\mu }$ pair 
\begin{eqnarray}
&&L=\sum f_{abc}F^{a.\mu \nu }A_{\mu }^{b}A_{\nu }^{c},~f\text{ }%
antisymmetric  \label{first} \\
&&Q_{\mu }=2\sum f_{abc}u_{\mu }^{a}F^{b,\mu \nu }A_{\nu }^{c}  \notag
\end{eqnarray}%
As in all previous calculations operator products of free fields are always
Wick-ordered. One expects that the Lie algebra restriction for the $f$%
-couplings results from the imposition of second order $e$-independence.%
\begin{eqnarray}
&&\mathfrak{A}_{sym}~\mathfrak{=~}\mathfrak{A+A}((x,e)\longleftrightarrow
(x^{\prime },e^{\prime })),~\mathfrak{A}=d_{e}T_{0}LL^{\prime }-\partial
^{\mu }T_{0}Q_{\mu }L^{\prime }=  \label{an} \\
&=&2\sum f_{abc}u^{b}A_{\nu }^{c}\{f_{def}\delta _{\kappa ,\lambda }^{\nu
}\delta ^{ad}A^{\prime d,\kappa }A^{\prime e,\lambda }+f_{daf}\delta
_{\lambda }^{\nu }F^{\prime d,\kappa \lambda }A_{\kappa }^{\prime
f}+f_{dea}\delta _{\kappa }^{\nu }F^{\prime d,\kappa \lambda }A_{\lambda
}^{e\prime }\}  \notag
\end{eqnarray}%
Here the deltas are the singular parts (s.p.) which result from derivatives
applied to $T_{0}~$propagators 
\begin{eqnarray}
\delta _{\kappa ,\lambda }^{\nu }(\xi ) &=&s.p.\partial _{\mu
}^{{}}\left\langle T_{o}F^{b,\mu \nu }(x)F_{d,\kappa \lambda }(x^{\prime
})\right\rangle =-i(g^{\nu \lambda }\partial ^{\prime \kappa }-g^{\nu \kappa
}\partial ^{\prime \lambda })\delta (\xi ))\delta ^{bd} \\
\delta _{\lambda }^{\nu }(\xi ) &=&s.p.\partial _{\mu }\left\langle
T_{0}F^{b,\mu \nu }(x)A_{a\lambda }(x^{\prime })\right\rangle =-ig_{\lambda
}^{\nu }\delta (\xi )-ie^{\prime \nu }\partial _{\lambda }^{\prime
}\int_{0}^{\infty }ds\delta (\xi -se^{\prime })  \notag
\end{eqnarray}

The second line contains a contribution from string crossing; the $s$%
-integral results from the Fourier transformation of the integrands (\ref%
{sdel}) 
\begin{equation}
\frac{1}{p^{2}-m^{2}}\frac{e^{\prime \nu }p_{\lambda }}{pe^{\prime
}-i\varepsilon }
\end{equation}%
in the $\left\langle TFA^{\prime }\right\rangle $ propagators. In writing
the first line we followed Scharf (page 113 in \cite{Scharf}) by using the
freedom of a normalization term (according to the Epstein-Glaser scaling
rules) in the various 2-derivative contributions to the time-ordered
two-point functions in $\left\langle TFF^{\prime }\right\rangle $ e.g.%
\begin{equation}
\partial ^{\mu }\partial ^{\nu }D(x-x^{\prime })\rightarrow \partial ^{\mu
}\partial ^{\nu }D(x-x^{\prime })+\alpha g^{\mu \nu }\delta (x-x^{\prime })
\end{equation}%
At the end of the calculation the remaining anomaly must of course be
independent of $\alpha .$

Symmetrizing in order to obtain $\mathfrak{A}_{sym\text{ }}$one notices that
for$~e=$ $e^{\prime }$ the string-local delta contributions cancel and one
arrives at 
\begin{eqnarray}
&&\mathfrak{A}_{sym}=f_{abc}u_{a}A_{c,\nu }[f_{bef}A_{e,\sigma }\partial
^{\sigma }A_{f}^{\nu }+f_{dbf}A_{d,\rho }\partial ^{\nu }A_{f}^{\rho
}]2\delta (x-x^{\prime }) \\
&&+f_{abc}f_{deb}[(\alpha +1)(\partial ^{\sigma }u_{a}A_{c,\nu
}+u_{a}\partial ^{\sigma }A_{c,\nu })A_{d}^{\nu }A_{e,\sigma }+  \notag \\
&&+(\alpha -1)u_{a}A_{c,\nu }(\partial ^{\sigma }A_{d}^{\nu }A_{e,\sigma
}+A_{d}^{\nu }A_{e,\sigma }]\delta (x-x^{\prime })  \notag
\end{eqnarray}%
The cancellation of the anomalies leads to $\alpha =1$ and the term $%
\partial ^{\sigma }u_{a}A_{c,\nu }A_{d}^{\nu }A_{e,\sigma }~$has the form 
\begin{equation}
d_{e}N_{2},~~N_{2}=\frac{1}{4}f_{abc}f_{dec}A_{a,\mu }A_{d}^{\mu }A_{b,\nu
}A_{e}^{\nu }\delta (x-x^{\prime })
\end{equation}%
$~$the The validity of the Jacobi identity is then a consequence of the
remaining cancellation.

Another more systematic bookkeeping (following the logic of $\mathfrak{A}%
_{sym}~$in section 3) would consist in converting derivatives of delta
functions $\partial _{\mu }\delta ...$into $\partial _{\mu }(\delta
...)-\delta \partial _{\mu }(...)$\footnote{%
This was pointed out to me by Jens Mund.}.$~$But here we followed Scharf in
order to emphasize the formal proximity to the gauge formalism despite the
conceptual differences between CGI and SLF.

Actually this derivation of the Lie-algebra structure of the second order
interaction density involved a bit of cheating since the operators $Q_{\mu
},~$which was used in intermediate steps, has no zero mass limit. Such
perturbative infrared divergencies are a warning against calculating
directly in the massless theory, instead of constructing it from the $%
m\rightarrow 0$ limit of massive correlation functions.

The analogous calculations in the massive model avoids these problems. But
as a consequence of the appearance of a compensatory
renormalization-preserving $H$-field it is more involved. The verification
of its Lie structure will be left to a joint publication with Jens Mund.

\section{Resum\'{e} and outlook}

The SLF setting extends the Hilbert space description of point-local $s<1$
interaction to $s\geq 1~$string-local interactions in which only the local
observables remain point-local. Since it is a Hilbert space formalism, the
powerful nonperturbative tools of functional analysis and operator algebras
are available. Gauge theory keeps the simpler point-local formalism for $s=1$
at the prize of loosing off-shell unitarity. Since the functional and
operator algebra tools are not available in Krein space there is no
nonperturbative control. The physical content is restricted to combinatorial
perturbative constructions; conceptual and mathematical aspects concerning
the relation between particles and fields remain outside its range.

The string-local fields $\Psi (x,e)$ are Wightman fields in both variables;
they have to be smeared with Schwartz testfunctions both in Minkowski
spacetime and in the de Sitter space of space-like directions $e$. The
differential geometry and the associated differential form calculus of $%
d=1+2 $ dimensional de Sitter space is a spacetime analogue of the abstract
BRST cohomology of gauge theory defined in terms of the nilpotent $\mathfrak{%
s}~$operation (the ghost formalism).

The continued validity of the appropriately adjusted LSZ or Haag-Ruelle
scattering theory \cite{Haag} \cite{Bu-Fr2} in the presence of a mass gap
permits to identify the Hilbert space with a Wigner-Fock space in which
scattering theory relates the Wigner particles to the interacting fields.
For interactions involving massless vector mesons this field-particle
relation breaks down and the S-matrix is lost; the fields are no longer
connected with Wigner particles. In the perturbative SLF Hilbert space
setting this manifests itself right from the beginning. A first order
interaction $L~$cannot be defined since a scalar string-local $\phi $ escort
field, and hence also $V_{\mu }$ and $Q_{\mu },~$have no massless limit. In
the following we will argue that this loss is actually an asset since it
points into new directions for solving problems which are outside the scope
of gauge theory.

The loss of the interaction density and a unitary S-matrix in the massless
limit is not the result of a short-coming of the formalism, but it rather
points to a fundamental physical change. Of all unsolved problems of QFT,
the ones hidden behind the infrared divergencies caused by massless vector
potentials have remained the hardest. Though particle physicists have use
recipes as e.g. photon-inclusive cross sections in QED and string-bridged
quark-antiquark pairs in order to describe hadronic jets, there has been no
spacetime understanding of these phenomena.

The message offered by the new SLF setting is that one should not address
problems of perturbative interacting zero mass vector mesons directly, but
rather treat them as limiting cases of their massive counterpart. The first
step consists in computing the correlations functions in the massive theory.
The Hilbert space positivity remains encoded in the Wightman positivity of
the vacuum expectation values in the massless limit \cite{St-Wi}. The
massless operator theory and its Hilbert space can be reconstructed from the
massless limit of these expectation values; in this way one avoids the
futile attempt of trying to understand its Hilbert space in terms of
particle concepts of the massive theory.

Such problems are outside the conceptual range of the BRST gauge formalism;
the locality of fields in Krein space is not that of the physical Einstein
causality even if it looks the same. In contrast to the Hilbert space
setting which, as emphasized before, signals the problems of zero mass
limits already in first order perturbation theory, gauge theory masks them.
The BRST formalism continues to work in the massless limit, but its physical
range is limited to gauge invariant local observables and prescriptions for
photon-inclusive cross-sections (whose gauge invariance is not obvious). The
success of the gauge theoretical description of the Standard Model is based
on the perturbative construction of the S-matrix for interactions massive
vector mesons; a deeper understanding of the physical content hidden behind
infrared divergencies requires the string-local setting; here the Hilbert
space positivity (off-shell unitarity) is indispensable.

In the present work the calculations within the SLF Hilbert space setting
were limited to the on-shell S-matrix. The important task of extending the
SLF setting to the construction of interacting string-local fields
(off-shell unitarity) will be addressed in future publications. The
remainder of this concluding section present an outlook about what one may
expect from such an extension for a better understanding of the infrared
problems.

The present picture based on perturbative calculations in gauge theory is
that, although the perturbative scattering amplitudes are logarithmically
infrared-divergent, the photon-inclusive cross section remains finite and
gauge invariant \cite{YFS}. It is believed that the logarithmic on-shell
divergencies for scattering of charge-carrying particles result from a
perturbative expansion of the "infraparticle structure". The latter is based
on the idea that the interaction with the infrared photons "dissolves" the
mass-shell of charged particles by converting the mass-shell poles into a
milder cut-like singularity with a coupling-dependent power behavior. Such a
"softened mass shell singularity" would be too weak to counteract the
dissipation of wave packets in the large time limits in LSZ scattering
theory which then accounts for the vanishing of the scattering amplitude.
The logarithmic divergencies of the perturbative scattering amplitude result
from the perturbative expansion of the softened mass shell.

In order to reconcile the vanishing scattering amplitudes with the
photon-inclusive cross section one again replaces the photons by massive
vector mesons of a small mass $m$. The number $N(m)$ of contributing vector
mesons below a given invariant energy increases with $m\rightarrow 0,$ but
the inclusive cross section remains nontrivial while the individual
amplitudes approach zero \cite{YFS}. In the SLF setting all objects which
enter this argument are physical.

The change of the mass-shell singularity in the infraparticle picture is
supported by soluble two-dimensional models of which the simplest was
already constructed in the 60s \cite{Fort}. The solution $\psi (x)=\psi
_{0}(x)\exp ig\varphi $ of the two-dimensional derivative coupling $g\bar{%
\psi}\gamma _{\mu }\psi \partial ^{\mu }\varphi $ provides an example in
which the mass shell of $\psi _{0}(x)$ changes into a $g$-dependent power
singularity in the massless limit of the scalar field $\varphi $. The
perturbative expansion of the exponential function before taking the
massless limit leads to the logarithmic infrared divergencies. The
simplicity of the model permits to study infrared behavior without dynamical
complications (see the remarks at the end of section 3).

The SLF Hilbert space setting is expected to lead to an understanding of
infrared phenomena in terms of large distance properties of string-local
charge-carrying fields. For this one would have to extend the S-matrix
formalism to correlation functions of string-local fields and define the
massless theory in terms of the massless limit of corresponding massive
correlation. The important role of the differential calculus on de Sitter
space of space-like directions is to enable the implementation of the
independence on \textit{inner string directions}. In case of the S-matrix
all $e^{\prime }s$ are inner, whereas in the extension to vacuum expectation
values one has to distinguish between the string directions of those fields
whose correlation functions one wants to calculate and$\ $the inner $%
e^{\prime }s$ of the perturbative internal propagators.

This is reminiscent of gauge invariance, except that the BRST $\mathfrak{s}$%
-operation has no connection with spacetime and cannot distinguish between
"inner and outer" gauge invariance. Without the local nature of the $d$
calculus, in which the $d_{e}$ differentials acts on individual fluctuating $%
e$ directions, it would not be possible to differentiate between outer and
inner directional fluctuations. Needless to add that string-localization
bears no relation to String Theory which, despite its name and inspite of
many attempts of its defenders to connect it to QCD strings, bears no
relation to localization in spacetime.

There is an important structural difference between interacting strings in
the presence of mass gaps and massless strings. Whereas in the former case
the correlation functions of string-local fields decrease exponentially for
large space-like separations (linked cluster property) \cite{Bu-Fr2} and the
string-directions can be freely changed by Lorentz transformation, the
strings of charge-carrying particles in QED are "rigid". This leads to the
spontaneously breaking of Lorentz invariance in charged sectors \cite{Froe}.
Needless to add that the correct QFT analogue of long range interactions
(Coulomb potentials) in quantum mechanics are string-local charge-carrying
fields in the Hilbert space description of QED.\ 

A more radical change is expected to occur in the massless limit of massive 
\textit{selfinteracting} vector mesons. As explained before, the logarithmic
divergent perturbative scattering amplitudes of charge-carrying particles
are viewed as resulting from an illegitimate interchange of the limit $%
m\rightarrow 0$ with the perturbative expansion in a situation in which the
non-perturbative limit vanishes. This suggests to view logarithmic
divergencies in off-shell correlation functions of QCD containing gluon and
quark fields as signaling confinement. In this case the only surviving
vacuum expectation values would be those of point-local composites
(gluonium, hadrons) and string-bridged compact localized $q-\bar{q}$ pairs.
A situation in which the basic fields (in terms of which one defines
interactions) vanish cannot be directly described within the known
formulation of QFT; one needs to define such interactions as massless limits.

This picture receives additional support by noticing a significant
structural difference between QED and QCD strings. String-local vector
potentials in QED are line integrals over \textit{observable} field
strengths and hence can be viewed as \textit{global limits of local
observables}. But interacting string-local fields in QCD cannot be
represented in this way; their localization is inherently non-compact and
their appearance in correlation functions would cause problems with
causality. Confinement, in the sense of vanishing of all vacuum expectation
values containing such inherent noncompact strings, avoids such causality
problems.

It has been shown that perturbative \textit{QCD correlation functions in
covariant gauges remain finite} \cite{Hollands}. But the gauge which
formally corresponds to the string-local Hilbert space formulation is the
non-covariant axial gauge. It was abandoned a long time ago because it leads
to uncontrollable (ultraviolet mixed with infrared) divergencies.

But what seems to be a curse in gauge theory turns out to be a blessing in
the SLF Hilbert space setting. The latter turns the ill-defined noncovariant
axial gauge parameter in Krein space into fluctuating directions of
covariant string-local fields which act in Hilbert space. In this way the
global on-shell unitarity of gauge theory is extended to local correlation
functions of fields. This extension is of special importance in zero mass
limits when the field-particle relation is lost. The SLF Hilbert space
setting permits to address structural changes in massless limits which
remain outside the reach of gauge theory, as they occur in the infraparticle
structure of QED and the expected QCD confinement.

To achieve this one must extend the SLF of the present work to the
perturbative calculation of correlation functions for massive string-local
fields. Only in this way one will be able to understand phenomena as
confinement in which the fundamental fields disappear in the massless limit
and only their imprint on their composites (e.g. hadrons, and gluonium) and
the disintegration of string-bridged $q-\bar{q}$ pairs remains. Our present
understanding of QFT permits no description of fields which, although not
present in the formalism, yet assert their presence in "what they leave
behind". The only known way to deal with such a situation is to view it as
the massless limit of a massive model in a (necessarily string-local)
Hilbert space setting.

\textbf{Acknowledgements} The interest in the construction of string-local
field and the study of their properties dates back to the 2005 joint work
with Jens Mund and Jakob Yngvason. Afterwords Jens and I became interested
in the use these ideas for a reformulation of gauge theory. Whereas he
started to look into the more ambitious project to extend the causal
Epstein-Glaser setting from point-local to string-local fields, my interest
was to relate the new ideas with the old ideas around the Schwinger-Swieca
charge screening with which I had some familiarity from my collaboration
with J. A. Swieca during the 70s. So Jens and I decided to temporarily
follow these interests separately in order to afterwards continue the shared
project. I am deeply indebted to him for many discussions and for making his
manuscript available to me. I am also indebted to Jos\'{e} M. Gracia-Bondia
for helpful comments.

Last not least I thank Raymond Stora for his critical sympathy and helpful
suggestions with which he accompanied the development of this project for
more than a year.

\end{document}